\documentclass[aps,a4paper,twocolumn,showpacs]{revtex4}

\usepackage{graphicx}

\begin{document}

\newcommand{\be}   {\begin{equation}}
\newcommand{\ee}   {\end{equation}}
\newcommand{\ba}   {\begin{eqnarray}}
\newcommand{\ea}   {\end{eqnarray}}
\newcommand{\tr}   {{\rm tr}}

\title{Entangling power of baker's map: Role of symmetries}

\author{R\^omulo F. Abreu}
\email{romulof@cbpf.br}

\author{Ra\'ul O. Vallejos}
\email{vallejos@cbpf.br}
\homepage{http://www.cbpf.br/~vallejos}
\affiliation{ Centro Brasileiro de Pesquisas F\'{\i}sicas (CBPF), \\
              Rua Dr.~Xavier Sigaud 150, 
              22290-180 Rio de Janeiro, Brazil}

\date{\today}

\begin{abstract}
The quantum baker map possesses two symmetries: 
a canonical ``spatial" symmetry, and a time-reversal symmetry. 
We show that, even when these features are taken into account,  
the asymptotic entangling power of the baker's map 
does not always agree with the predictions of random matrix 
theory.
We have verified that the dimension of the Hilbert space is the
crucial parameter which determines whether the entangling 
properties of the baker are universal or not.
For power-of-two dimensions, i.e., qubit systems, an anomalous
entangling power is observed; otherwise the behavior of the
baker is consistent with random matrix theories.
We also derive a general formula that relates the asymptotic 
entangling power of an arbitrary unitary with properties of 
its reduced eigenvectors.

\end{abstract}



\maketitle

\section{Introduction}

The baker's map was quantized in 1987 by Balazs and Voros 
\cite{balazs87} and soon became a very useful toy model for 
investigating quantum-classical correspondence issues in 
closed chaotic systems, like the scarring phenomenon,
Gutwiller trace formula and the long-time validity of 
semiclassical approximations (see, e.g., 
\cite{balazs89, saraceno90, ozorio91, saraceno94, dittes94, 
heller96}).

Later on, the quantum baker appeared in a variety of problems 
of Quantum Information, Quantum Computation and Quantum Open 
Systems.
Schack noted that the quantum baker could be efficiently 
realized in terms of quantum gates \cite{schack98}.
A three qubit Nuclear Magnetic Resonance experiment was 
proposed \cite{brun99} and then implemented (with some 
simplifications) \cite{weinstein02}.
On the theoretical side, Schack and Caves \cite{schack00} showed 
that the quantum baker of Balazs and Voros' can be seen as a shift 
on a string of quantum bits --in full analogy with the classical 
case-- and exhibited a family of alternative quantizations.
This family of bakers was the subject of several studies
\cite{soklakov00,tracy02,scott03}. 
Decoherent variants of the baker map have been constructed 
by including mechanisms of dissipation and/or diffusion 
\cite{lozinski02, soklakov02, bianucci02}.

The ability of the baker family to generate entanglement
was studied by Scott and Caves \cite{scott03}. 
They concluded that 
``the quantum baker's maps are, in general, good at creating 
multipartite  entanglement amongst the qubits. 
It was found however, that some quantum baker's maps can, 
on average, entangle better than others, and that all quantum 
baker's maps fall somewhat short of generating the levels of 
entanglement expected in random states. 
This might be related to the fact that spatial symmetries in
the baker's map allow deviations from the predictions of random 
matrix theory." \cite{scott03}

The purpose of the present paper is to demonstrate that the 
spatial symmetry is not to blame for the reduced entangling 
power of the baker. 
Two numerical complementary proofs will be presented. 
First we check that if the symmetry is removed from the baker
(by block diagonalization), the resulting desymmetrized bakers 
produce the same levels of entanglement as the original one.
Second we show that the entangling power of an ensemble of 
``spatially-symmetric" unitary operators is not significantly 
different from that of the CUE ensemble \cite{mehta04} of random 
operators, i.e., imposing symmetry does not reduce the entangling 
power. 
We complete the analysis by verifying the dimension of the Hilbert 
space is indeed the crucial parameter which determines whether 
the entangling properties of the baker are universal or not.
For qubit systems, i.e., power-of-two dimensions, an anomalous
entangling power is observed; otherwise the behavior of the
baker is consistent with random matrix theories.

The background of this contribution is the wider problem of 
understanding what makes a unitary operation a good entangler.
There is not a definitive answer to this question yet, but 
some partial results have been obtained recently 
(see, e.g., \cite{gorin03,demkowicz04,bandy04,weinstein04}).

Section \ref{sec2} presents the measure we use for quantifying 
the entangling power of a unitary operation.  
The quantum maps to be considered are introduced in section 
\ref{sec3}, and their symmetries analyzed.
Sections \ref{sec4} and \ref{sec5} contain the numerical analysis 
of the influence of the symmetries on the entangling power of the 
baker.
The general relation between asymptotic entangling power and 
eigenvectors is discussed in \ref{sec6}.
Concluding remarks are presented in section \ref{sec7}.

\section{Entangling power}
\label{sec2}

We will only consider the case of bipartite entanglement of pure 
states. 

A full system, with Hilbert space of dimension $d = d_A \times d_B$, 
is partitioned into two subsystems, $A$ and $B$, with dimensions 
$d_A$ and $d_B$, respectively, and such that the full space $\cal{H}$ 
has the structure of a tensor product, 
${\cal H}={\cal H}_A \otimes {\cal H}_B$.
A usual definition of entangling power, $e_p(U)$, of a unitary 
operator $U$ defined on $\cal{H}$ relies on a vector 
entanglement measure $E$, and on a suitable average over initial 
states \cite{zanardi00}:
\be
\label{epofu}
e_p(U) = \left \langle 
      E \left( U | \psi_A \rangle \otimes | \psi_B \rangle \right)
         \right \rangle_{|\psi_A \rangle , |\psi_B \rangle} \;.
\ee
Thus, $e_p(U)$ says how much entanglement $U$ produces, on 
average, when acting on a set of non-entangled states. 
Here we take the measure $E$ to be the linear entropy of the 
reduced density matrix: 
Let $|\psi \rangle $ be a (pure) separable state of the full 
system, corresponding to the density matrix
$\rho = |\psi \rangle \langle \psi |$.
In general, after application of $U$, the new density matrix 
$\rho^\prime = U \rho U^\dagger$ 
will not correspond to a separable state any more. 
This will manifest in a positive linear entropy of the reduced 
density matrices 
\be
S_L \equiv 1 - \tr \left( {\rho}^\prime_A \right)^2 
       =   1 - \tr \left( {\rho}^\prime_B \right)^2
       >   0   \; ,  
\ee 
where
${\rho}^\prime_A = \tr_B \rho^\prime$  and  
${\rho}^\prime_B = \tr_A \rho^\prime$  
\cite{nielsen01}.

For the average over product states, indicated by 
$\langle \ldots \rangle_{|\psi_A \rangle , |\psi_B \rangle}$
in (\ref{epofu}), we choose the unitarily invariant measures in 
both ${\cal H}_A$ and ${\cal H}_B$ \cite{wootters90,zanardi00}. 
That is, the components of $|\psi_A \rangle$ and $|\psi_B \rangle$ 
have the same distribution as the columns of CUE matrices 
of dimension $d_A \times d_A $ and $d_B \times d_B$, respectively 
\cite{emerson03}.

Among various possible definitions of the entangling strength 
of a unitary \cite{zanardi00,zanardi01,wang02,nielsen03}, we 
chose Eq.~(\ref{epofu}) mainly for the purpose of comparing our 
results with those in Ref.~\cite{scott03}, where that definition 
was adopted.
The use of $S_L$ instead of the more natural von Neumann entropy 
not only does not lead to qualitative differences \cite{scott03}
but has the essential advantage of allowing analytical calculations, 
which will be important for understanding and extending our 
results.

\section{Quantum maps, symmetries}
\label{sec3}

Following Schack and Caves \cite{schack00}, we write the unitary 
operator for the quantum baker on $N$ qubits as ($d=2^N$) 
\be 
B_d = G_d \left( \openone_2 \otimes G^{-1}_{d/2} \right) \; ,
\label{baker}
\ee 
where $G_d$ is the antiperiodic quantum Fourier transform on $N$ 
qubits, $\openone_2$ is the unit operator for the first qubit, 
and $G^{-1}_{d/2}$ the inverse (antiperiodic) Fourier transform 
on the remaining $N-1$ qubits. 
We use the standard ordering for the computational basis: 
If $|j \rangle$ is a tensor product of individual qubit
basis states $|\epsilon_i\rangle$, with $\epsilon_i=0,1$, i.e., 
\be
|j \rangle = 
|\epsilon_1 \rangle \otimes | \epsilon_2 \rangle \otimes \ldots
|\epsilon_N \rangle \; ,
\ee
then $j$ is given by the binary expansion
\be
j = \sum_{i=1}^N \epsilon_i 2^{N-i} 
  \equiv \epsilon_1 \ldots \epsilon_N  \; ,
\ee
$0 \le j \le d-1$.

The definition (\ref{baker}) becomes equivalent to the baker 
of Balazs-Voros and Saraceno when one identifies the computational 
basis states $|j \rangle$ with ``position" eigenstates 
$|q_j \rangle$ having eigenvalues 
$q_j= 0.\epsilon_1 \ldots \epsilon_N 1$. 
In any case, the matrix representation of the baker is
\be
\| B_d \| = \| G_d \|
                \left( \begin{array}{cc}
          \| G^{-1}_{d/2} \| & 0 \\
                           0 & \| G^{-1}_{d/2} \|
                       \end{array}
                \right) \; ~,
\label{bakermatrix}
\ee
where $\| G_d \|$ is the inverse Fourier {\em matrix} 
\cite{balazs89,saraceno90}.

The baker map has a time-reversal symmetry, corresponding to the 
Fourier transform followed by complex conjugation in the
computational basis:
\be
\| G^{-1}_d B_d G_d \|^\ast = \| B_d^{-1} \| \; ~.
\ee
The use of the antiperiodic Fourier transform makes the baker 
also reflection symmetric, in agreement with its classical 
counterpart \cite{saraceno90}. 
That is, if we define the reflection operator
\be
R_d |j \rangle = |d - 1 - j \rangle \; ,
\ee
then
\be
B_d R_d = R_d B_d \; .
\label{symmetry}
\ee
For a qubit system the reflection operator can be factored 
into a tensor product of $N$ single qubit reflections 
\be
R_d |j \rangle =  R_2 |\epsilon_1 \rangle \otimes 
                  R_2 |\epsilon_2 \rangle \otimes 
                  \ldots \otimes 
                  R_2 |\epsilon_N \rangle \; ,
\ee
where $R_2$ is just the negation operator (Pauli-$X$ gate).
The reflection symmetry of the baker, Eq.~(\ref{symmetry}), can 
be easily proved using the reflection symmetry of the antiperiodic 
Fourier transform, 
$R_d G_d = G_d R_d$,
and the factorization property 
$R_d = R_2 \otimes R_{d/2}$.

Given that the quantum baker is a unitary operator with a chaotic 
classical limit, one may expect that the iterative application of
the baker to random non-entangled states could produce states with 
levels of entanglement typical of random states. 
However, Scott and Caves verified that, in spite of being a good 
entangler, the baker generates states that are somewhat less 
entangled than random states. 
They suggested that this deviation might be due to the spatial 
symmetry $R_d$, which, as any unitary symmetry, is known to produce 
deviations from standard random matrix behavior. 

In the following we implement a simple test to decide if the spatial 
symmetry does really play a significant role in the reduction of the
entangling power of the baker.
Due to the symmetry $R_d$, with eigenvalues $\pm 1$, the baker can 
be cast into block-diagonal form: 
\be
\label{breduced}
\Lambda^\dagger \, B_d \, \Lambda = |0 \rangle \langle 0 | 
                   \otimes B^{(-)}_{d/2} +  
                  |1 \rangle \langle 1 | 
                   \otimes B^{(+)}_{d/2}   \; ~,
\ee
where $B^{(\pm)}_{d/2}$ are the symmetry-reduced baker maps,
and $\Lambda^\dagger$ is a unitary mapping of the computational basis 
to a $R_d$-symmetrical basis:
\be
\label{lambda}
\Lambda = \frac{1}{\sqrt{2}} 
    \left( \openone_d + i Y \otimes R_{d/2} \right) ,
\ee
with $Y$ the second Pauli matrix.

The maps $B^{(\pm)}_{d/2}$ have well known classical limits:
They correspond to conservative, piecewise linear versions 
of the Smale horseshoe \cite{cvitanovic88,saraceno96}.
Instead of working with the exact $B^{(\pm)}_{d/2}$, we prefer 
the simpler approximate expressions:
\ba  
    B^{(-)}_d     & \cong &   G_d \left( |0 \rangle \langle 0 | 
                                    \otimes G^{-1}_{d/2} +
                                    |1 \rangle \langle 1 | 
                                    \otimes G_{d/2}    
                             \right)  \nonumber \\ 
                  & \equiv & D_d \, , \label{dmap} \\
R_d B^{(+)}_d R_d & \cong  & G_d \left( |0 \rangle \langle 0 | 
                                    \otimes G^{-1}_{d/2} -
                                    |1 \rangle \langle 1 | 
                                    \otimes G_{d/2}    
                             \right)  \nonumber \\
                  & \equiv & D^\prime_d \, .  
\ea 
The operators on the RHS in the equations above, 
$D_d$ and $D^\prime_d$,
are both unitary, time-reversal symmetric, and do not have 
spatial symmetries. 
They share a same classical limit with the reduced bakers, 
but correspond to a slightly different quantization of the 
classical baker map \cite{saraceno96}. 
Instead of (\ref{breduced}), $D$ and $D^\prime$ satisfy
\be
\Lambda^\dagger \, {\overline B}_d \, \Lambda = 
     |0 \rangle \langle 0 | D_{d/2} +  
     |1 \rangle \langle 1 | R_d D_{d/2}^\prime R_d \; ~,
\ee
where ${\overline B}_d$ coincides with Schack and Caves' 
baker $\hat{B}_2$ \cite{schack00}.

\section{Comparing $B_d$ with $D_d$}
\label{sec4}

In order to assess the effect of symmetries in the production
of entanglement, it suffices to consider the non-symmetric 
operators $D_d$ or $D^\prime_d$, and compare with $B_d$.

\begin{figure}[htp]
\hspace{0.0cm}
\includegraphics[width=9cm]{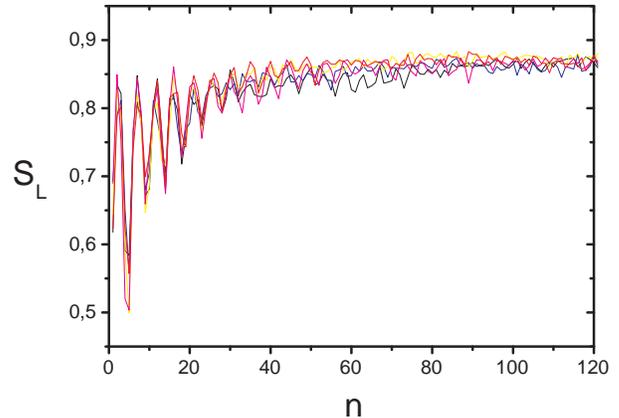}
\caption{(color online) 
Five initial pure product states were chosen randomly 
according to the CUE measure, and then evolved by applying the 
$D$-map $n$ times. At each time the linear entropy, $S_L(n)$,
is calculated. The system consists of eight qubits divided into
two subgroups of most/least significant four.}
\label{fig1}
\end{figure}

The iterative application of $B_d$ to an initial product state
typically makes the entropy grow quickly from zero to some
``equilibrium" value, which depends on the initial state.
After that, the entropy keeps fluctuating with small amplitude 
around the equilibrium. This is the behavior observed by Scott
and Caves for a variety of ways of partitioning the qubits 
\cite{scott03}. 
Our analises of the map $D_d$ [Eq.~(\ref{dmap})] verify the same 
qualitative features. 
As an example, we show in Fig.~\ref{fig1} some plots of entropy 
vs. time for a system of eight qubits split into two groups: 
the four most significant qubits on one side, the remaining 
least significant on the other.

From now on we focus on the asymptotic regime of large times, 
when the system has already relaxed to equilibrium. 
In this regime we expect the statistical properties of 
entangled states to be described by some random matrix model. 
The simplest ansatz associates evolved pure states with
random states of the same Hilbert space, chosen according to
the CUE measure. 
The average linear entropy of these states is given by
\be
\left \langle S_L \right \rangle_{\rm CUE} = 
\frac{(d_A-1)(d_B-1)}{d_A d_B+1} \; ;
\ee
analytical expressions for the second and third cumulants are 
also known (see \cite{scott03} and references therein).

\begin{figure}[htp]
\hspace{0.0cm}
\includegraphics[width=9cm]{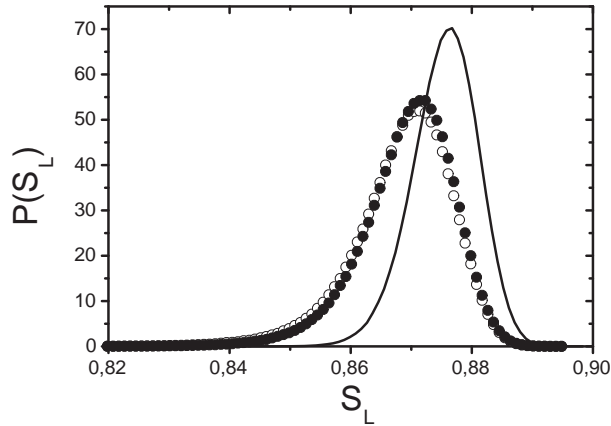}
\caption{Histograms of asymptotic linear entropies generated by 
the baker map (open circles) and by the map $D_d$ (full circles).
The line corresponds to the distribution of entropies for 
a large set of CUE random states calculated numerically.
In all cases the system consists of eight qubits divided into 
two subgroups of most/least significant four.}
\label{fig2}
\end{figure}

For a quantitative comparison between $D$-map and baker, 
we considered the same system as before, but this time we 
generated a set 
of $2 \times 10^6$ data by gathering values of $S_L(n)$ for 
$513 \le n \le 2512$ and $10^3$ random initial states.
These data, properly binned, are displayed in Fig.~\ref{fig2} 
together with analogous data for the baker. 
It can be immediately seen that both maps, baker and $D_d$, 
produce very similar distributions of entropies, both shifted 
to values lower than those of random states. 
The conclusion of this comparative simulation is that the
reflection symmetry is not the cause for the states generated
by the baker being less entangled, in average, that random
states (because $D$-map is non-symmetric and also shows a
reduced entangling power). 
However, absence of symmetry may be the explanation for a very 
small, though perceptible, increase of entangling power of the 
$D$-map as compared with the baker (see Fig.~\ref{fig2}). 
(Similar influence of symmetry in coupled tops was reported by 
Bandyopadhyay and Lakshminarayan \cite{bandy02}.)

Another simple complementary test, which also shows that the 
influence of the symmetries is very limited, consists of 
calculating the entangling power of a random map having the 
same symmetries as the baker, i.e., time-reversal and 
reflection symmetries. 

The problem of introducing symmetries in a random matrix 
model is well known, e.g., in the scattering approach to 
electronic transport through mesoscopic cavities 
\cite{gopar96,baranger96,zyczkowski97,aleiner00}: 
The unitary (scattering matrix) is brought to block diagonal 
form by choosing a basis with well defined symmetry, and 
each block is modeled by a circular ensemble. 
Using this recipe for the baker map we arrive at an ensemble 
of random matrices $\mathcal{B}$ with the structure
\be
\label{srmt}
\mathcal{B}  = 
     \Lambda 
        \left( |0 \rangle \langle 0 | \otimes W^{(1)} +  
               |1 \rangle \langle 1 | \otimes W^{(2)}  
        \right) 
     \Lambda^\dagger \; ,
\ee
where $W^{(1)}$ and $W^{(2)}$ are drawn independently from the
COE ensemble, appropriate for unitary maps with time
reversal symmetry \cite{mehta04}, and $\Lambda$ is defined in 
(\ref{lambda}). 

\begin{figure}[htp]
\hspace{0.0cm}
\includegraphics[width=9cm]{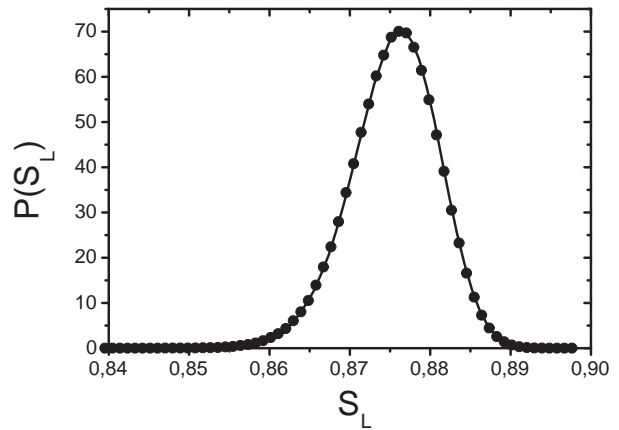}
\caption{Histograms of linear entropies generated by one
application of random maps belonging either to 
the symmetric ensemble of Eq.~(\ref{srmt}) (circles), or
to the CUE ensemble (line).
In the first case we applied 1000 symmetric maps to 1000 
random non-entangled states. 
The CUE data set is the same as in Fig.~\ref{fig2},
and so is the system of qubits.}
\label{fig3}
\end{figure}

In Fig.~\ref{fig3} we calculate the entangling power of the 
symmetric random maps defined above and compare with random maps 
having no symmetries at all, i.e., the CUE ensemble. 
The figure shows that differences between both ensembles are not 
significant. 

(CUE matrices were generated using the Hurwitz parameterization
\cite{pozniak98,forrester}. COE matrices were obtained simply 
by forming the products $V V^T$, with $V$ belonging to CUE 
\cite{mehta04}.)

%
%

\section{From qubits to arbitrary dimensions}
\label{sec5}

In the search for an alternative explanation for the anomalies 
observed, we recall that baker maps in spaces of power-of-two
dimensionality are known to exhibit peculiar properties. 
For instance, Balazs-Voros and Sano observed that bakers of dimension 
256 and 1024, respectively, display spectral statistics quite far 
from universal behavior \cite{balazs89, sano00}.
Though asymptotic entangling power is not a property determined 
by the eigenvalues, but by the eigenvectors (see below), the latter
may also be anomalous for the qubit case.
Thus, we now proceed to check if the precise value of Hilbert space 
dimension has a definite influence in the entangling properties of 
the baker.

\begin{figure}[htp]
\hspace{0.0cm}
\includegraphics[width=9cm]{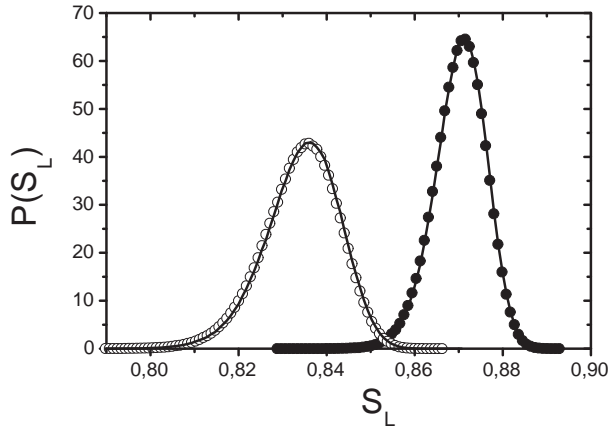}
\caption{
Histograms of asymptotic linear entropies generated by
two non-qubit baker maps with: 
(full circles) $d=238$, $d_A=14$, and $d_B=17$;
(open circles) $d=162$, $d_A= 9$, and $d_B=18$.
Data were collected by the same procedure used for
Fig.~\ref{fig2}.
Full lines are the corresponding CUE predictions.}
\label{fig4}
\end{figure}

Note that Eq.~(\ref{bakermatrix}) defines a quantum baker for any 
even dimension $d$. 
We shall consider, for instance, two systems with dimensions 
$d=238$ and $d=162$, partitioned as 
$238 =14 \times 17$ and $162 = 9 \times 18$.
The analyses of these cases is presented in Fig.~\ref{fig4}.
The histograms clearly show that, by avoiding power-of-two 
dimensions, we recover universal behavior. 
These results give additional support to the belief that qubit 
bakers (and $D$-maps) possess hidden symmetries, as happens 
with cat maps and with certain triangles of the hyperbolic 
plane \cite{bogomolny97}.

The case $d = 162 =2 \times 3^4$ was chosen to demonstrate that 
a power-of-three factor in Hilbert space dimension is not a 
source of anomaly for the baker. 
Presumably, power-of-three dimensions, i.e., qutrit systems, are 
anomalous for {\em ternary} bakers, like those considered by
Nonnenmacher and Zworsky \cite{nonnen05a,nonnen05b}.

%
%

\section{Relation with eigenvectors}
\label{sec6}

We have characterized the entangling abilities of a unitary
operator by the distribution of entropies it produces when 
acting iteratively on a set of non-entangled states.
But such an entropy distribution depends (weakly) on 
the number of iterations, even in the long-time regime.
For this reason, in the histograms of Figs. \ref{fig2} and 
\ref{fig4} we also included data corresponding to different 
times. 

Then it is natural to consider an average of the entropy 
over both initial states {\em and} time.
In this way we arrive at a quantity that characterizes 
unambiguously the {\em asymptotic entangling power} of an 
unitary, $e^\infty_p(U)$, defined by 
\be
\label{asymp}
e^\infty_p(U) \equiv 
         \lim_{K \to \infty} 
               \frac{1}{K} \sum_{k=1}^K e_p(U^k) \; ,
\ee
i.e., the time average of the entangling power of Eq.~(\ref{epofu}). 

In the case of the bakers and $D$-map considered in the previous
sections $e^\infty_p$ can be estimated as the mean value of the data set
used to generate each histogram in Figs. \ref{fig2} and \ref{fig4}.
However, extracting $e^\infty_p(U)$ from a finite data set 
introduces undesirable statistal errors. 
This could be avoided by implementing the averages in (\ref{asymp}) 
analitycally, followed by numerical evaluation of the resulting 
expression. 
We shall see in the following that this procedure leads to a relation 
between the asymptotic entangling power of an arbitrary unitary 
$U$ and its reduced eigenvectors. 
Such a relation is not only very interesting by itself, but
it will give us the possibility of checking the numerical 
simulations of previous sections.

We start by noting that if the average over initial states is removed 
from Eq.~(\ref{asymp}) one obtains the {\em asymptotic entropy} 
considered by Demkowicz-Dobrzanski and Kus \cite{demkowicz04}:
\ba
& & \hspace{-7ex} 
S^\infty_L (|\psi \rangle) =  1 - 
       \sum_i    \left| \langle e_i | \psi \rangle \right|^4
                        \tr_A \left( \rho_A^{ii} \right)^2  -   \nonumber \\
& &  - \sum_{i \neq j} 
                 \left| \langle e_i | \psi \rangle \right|^2
                 \left| \langle e_j | \psi \rangle \right|^2
                 \left[ \tr_A \left( \rho_A^{ii} \rho_A^{jj} 
                              \right) +  
                 \right.                                        \nonumber \\
& & \hspace{20ex}
               + \left. \tr_A \left( \rho_A^{ij} \rho_A^{ji} 
                              \right)   
                 \right]  ,  
\label{sinf}
\ea
which depends of the initial state $|\psi \rangle$.
In the expression above $|e_i \rangle$ stands for an
eigenvector of $U$, and $\rho_A^{ij}$ is the reduced operator
\be
\rho_A^{ij} \equiv \tr_B |e_i \rangle \langle e_j |   \; .
\ee
Of course, $\rho_A^{ii}$ is the reduced density matrix obtained
from the i-th eigenvector, to be denoted simply by $\rho_A^{i}$.

Formula (\ref{sinf}) is not of completely general validity. 
However, its derivation makes only the weak assumption that 
eigenvalues $\exp(i\phi_k)$ do not satisfy the commensurability 
relation
\be
\label{comm}
\phi_k - \phi_l + \phi_m - \phi_n = 0  \; ,
\ee
except for the trivial cases $k=l$ and $m=n$, or $k=n$ and $l=m$
(this condition is more general that merely requiring absence of 
degeneracies). 
We have verified numerically that the eigenvalues of the maps 
studied in this paper, even if anomalous in other sense, are not 
commensurable.
So, we can safely use (\ref{sinf}). 

Averaging (\ref{sinf}) over initial random product states we 
arrive at a formula for  $e^\infty_p(U)$ in terms of the 
eigenvectors of $U$.
The derivation is somewhat lengthy but simple.
It requires averaging products of the type 
\be
     c_{\alpha}       c_{\beta } 
     c_{\gamma}^\ast  c_{\delta}^\ast  \; ,
\ee
where $c_\alpha$ are the coefficients that arise from 
expanding the state $|\psi \rangle$ in the eigenbasis of $U$.
The coefficients $c_\alpha$ are distributed like the elements
in a column of a CUE matrix. 
The average above is one among others calculated by Mello some
time ago \cite{mello90}. 
We omit the details and just show the final result:
\ba
& &  e^\infty_p(U) =     
          \frac{d + 1 }{d'} - \frac{2}{d d'} \sum_{i} 
          \left[ \tr_A \left( \rho^{i}_A \right)^2
          \right]^2  -                         \nonumber \\
& &                   - \frac{1}{d d'}
                        \sum_{i \neq j }
          \left [ \tr_A \left( \rho^{i}_A \rho^{j}_A \right) + 
                  \tr_B \left( \rho^{i}_B \rho^{j}_B \right)
          \right]^2                            \; .
\label{einf}  
\ea
Here we used the abbreviations 
$d=d_A d_B$ and $d^\prime=(d_A+1)(d_B+1)$, together with the 
property
\be
\tr_A \left( \rho_A^{ij} \rho_A^{ji} \right)  =
\tr_B \left( \rho^{i}_B \rho^{j}_B \right)  \; ,
\ee
showing that (\ref{einf}) and (\ref{sinf}) are indeed 
invariant with respect to the swap of subsystems $A$ and $B$.

Equation~(\ref{einf}) is a useful formula which has absorbed 
the averages analitycally;
it expresses the asymptotic entangling power of a unitary as a 
function of a special combination of pairs of eigenvectors, 
\be
\tr_A \left( \rho^{i}_A \rho^{j}_A \right) + 
\tr_B \left( \rho^{i}_B \rho^{j}_B \right) \; ,
\ee
i.e., the symmetrized Hilbert-Schmidt scalar product of the 
reduced density matrices. 
For the maps considered here, we have checked that the calculation
of $e_p^\infty$ using either (\ref{einf}) or by straightforward
time and ensemble averages leads to consistent results; thus 
we verified the correctness of both procedures.

We remark that $e_p^\infty$ is not directly related to the 
eigenvector entropies 
\be
1-\tr_A \left( \rho^{i}_A  \right)^2 \; ,
\ee
even though, bounds relating asymptotic entangling power and 
eigenvector average entropies can be obtained by the use of
Cauchy-Schwartz inequality \cite{demkowicz04}. 
(One must remember that in some cases eigenvector entanglement 
may give a wrong estimatimation of $e_p^\infty$ 
\cite{demkowicz04, bandy05}.)

\section{Concluding remarks}
\label{sec7}

We demonstrated that the deviations from universal behavior
reported by Scott and Caves \cite{scott03} are not due to spatial
or time-reversal symmetries.
Instead, the anomalous entangling power of the qubit baker
originates from specificities associated to the dimension of
the Hilbert space being a power of two.
When other dimensions are considered, a behavior consistent
with random matrix theory is recovered.
Presumably qubit bakers possess symmetries of number theoretic
origin, i.e., with no classical analogues 
(``pseudo-symmetries" \cite{bogomolny97}).
All the members of the Schack-Caves family \cite{schack00} 
(of which the baker considered here is a special case) suffer, 
to different extent, from a reduction of the entangling 
power \cite{scott03}.
It is tempting to speculate that those differences may be 
related to each member having a different number of 
pseudo-symmetries.

\begin{acknowledgments}
%
We thank A. M. Ozorio de Almeida and M. Saraceno for many 
interesting comments.
Partial financial support from 
CNPq, CAPES, and The Millennium Institute for Quantum Information 
is gratefully acknowledged. 
\end{acknowledgments}


\end{document}